\documentclass[a4paper,11pt]{article}
\usepackage{jinstpub} 
\usepackage{url}

\newcommand{\ndbd}{\ensuremath{0\nu \beta\beta}}

\title{Vibrational sensing at mK temperatures in dry dilution refrigerators using commercial accelerometers for diverse fundamental physics applications}

\keywords{Cryogenic detectors, Cryocoolers, Vibration analysis, Instrumental noise}

\arxivnumber{2601.08817}

\author[d]{N.~Brace}
\author[a]{A.~D'Addabbo}
\author[a,b]{S.~D'Eramo}
\author[a]{S.H.~Fu}
\author[c]{M.T.~Hurst}
\author[d]{T.~O'Donnell}
\author[e]{S.~Petti\footnote{currently at Dip. di Fisica Università di Torino, Torino (TO) I-10125, Italy}}
\author[c]{V.~Sharma}
\author[c]{P.~T.~Surukuchi\footnote{Corresponding author.}}
\author[d]{A.~Torres}
\author[f]{K.~J.~Vetter}
\author[d]{C.~Wengappuliarachchige}

\affiliation[a]{INFN -- Laboratori Nazionali del Gran Sasso, Assergi (L'Aquila) I-67100, Italy }
\affiliation[b]{DIIIE, Università degli Studi dell’Aquila, Monteluco di Roio, L’Aquila 67100, Italy}
\affiliation[c]{Department of Physics and Astronomy, University of Pittsburgh, Pittsburgh, PA 15260, USA }
\affiliation[d]{Center for Neutrino Physics, Virginia Polytechnic Institute and State University, Blacksburg, Virginia 24061, USA}
\affiliation[e]{INFN -- TIFPA, c/o Dip. di Fisica Università di Trento, Povo (TN), I-38123, Italy}
\affiliation[f]{Laboratory for Nuclear Science, Massachusetts Institute of Technology Cambridge, MA 02139, U.S.A}

\emailAdd{surukuchi@pitt.edu}
\abstract{
This article presents an evaluation of off-the-shelf commercial accelerometers at the mixing chamber stage of a cryogen-free dilution refrigerator at temperatures down to 8 mK.
In addition, we present results of radioassay of accelerometers using a high purity germanium detector counting setup.
Cryogen-free dilution refrigerators using pulse-tube cryocoolers (PTs) --- due to recent advances in their cooling capacity, long-term stability, and operational costs --- have become ubiquitous tools in a wide range of fields ranging from experimental particle physics to quantum information sciences.
However, vibrations induced by PTs can negatively impact the experimental payload in these applications.
This work demonstrates that commercially available accelerometers can not only measure vibrations at millikelvin cryogenic temperatures but also pave the way for continuous, in situ, real-time vibration monitoring of dry dilution refrigerators.
This monitoring capability facilitates applications such as real-time denoising for vibration-sensitive experiments, thereby enabling ongoing noise assessment and mitigation.
}

\begin{document}  
\maketitle
\flushbottom

\section{Introduction and Motivation}
\label{sec:intro}

Cryogenic calorimetry is among the most sensitive technologies for a wide range of fundamental physics applications, including neutrinoless double beta decay searches \ndbd~\cite{CUORE:2021ctv, CUPID:2025avs, Auguste:2024xrg, AMoRE:2024loj}, coherent elastic neutrino–nucleus scattering~\cite{Ricochet:2021rjo,NUCLEUS:2017htt}, direct neutrino mass measurements~\cite{ECHO,HOLMES}, and dark matter experiments~\cite{SuperCDMS:2016wui, CRESST:2019jnq, EDELWEISS:2020fxc}.
These capabilities rely on stable operation of large detector volumes at millikelvin~(mK) temperatures.
Dilution refrigerators operate on the principle of $^{3}$He-$^{4}$He mixing~\cite{Enss:2005md} and 
can be designed to reliably cool large mass payloads down to below 10mK~\cite{cousins1999advanced}.

Wet dilution refrigerators, employing liquid helium for pre-cooling, are well-established and have been used extensively for decades. 
Dry dilution refrigerators~(DDR), using mechanical cryogen-free pre-cooling, are increasingly replacing wet refrigerators due to their growing cooling powers, long-term operational stability~\cite{CUORE:2024ikf}, and reduced reliance on volatile helium supplies.

The absence of moving parts at the cold stage makes PTs a lower-vibration alternative to other cryocooler technologies such as GM tubes for pre-cooling DDRs, and have been widely adopted in various applications, ranging from fundamental physics to the rapidly expanding field of quantum computing~\cite{Hollister:2024plk}.
However, pressure wave-induced vibrations can still propagate down to the experimental volume of the DDRs, thereby limiting their capabilities.
In cryogenic calorimeters, these vibrations manifest as reduced energy resolution~\cite{Ricochet:2025bpk}, elevated detector thresholds, and reduced efficiency~\cite{CUORE:2025aro}.
Furthermore, they present a growing challenge for the usage of DDRs in quantum sensing and quantum computing applications~\cite{QIS,Binney:2026wpj}.

To address these limitations, vibration reduction through passive decoupling of the stages of DDRs has emerged as an important area of research.
Approaches, such as springs~\cite{spring_Lee:2017hya,spring_NUCLEUS_decoupling,spring_Kellermann:2024dwt,spring_CUPID:2023hxo,spring_Maisonobe:2018tbq, CUORE:2015thw}, bellows~\cite{Olivieri:2017lqz}, and foam~\cite{Ricochet:2025bpk}, have been explored with varying success. 
The performance of the decoupling systems can be evaluated by deploying vibration sensors, such as accelerometers, on their room temperature flange~(300 K plate)~\cite{Olivieri:2017lqz}.
But this approach assumes vibration characteristics are consistent between the 300 K plate and the experimental volume, which is not universally applicable.
Geophones have also been employed~\cite{vanHaan:2013vjg, 10.1063/5.0053381, Lee:2018pxi, deWit:2018itv,Blair_1991} for vibration sensing across room-temperature and cryogenic environments, down to 3 K. 
However, low-frequency geophones are typically large in volume, posing challenges for space-constrained, thermally-restrictive, or radiation-sensitive applications.
Therefore, it is desirable to deploy low-mass vibration sensors, as witness channels, close to the experimental volume.

Furthermore, as experiments become increasingly sensitive to low vibration levels, not only from PTs but also other ambient sources such as microseisms~\cite{CUORE:2025aro}, the need for continuous vibration detection and monitoring with high sensitivity is increasingly critical.
A particularly compelling example is the CUORE experiment, in which vibrations are measured using several witness channels --- including accelerometers, seismometers, and microphones --- installed around the cryostat. 
These measurements are then correlated to effectively reduce noise~\cite{Vetter:2023fas} in the calorimeters suspended from the mixing chamber~(MC) plate. 
To further improve sensitivity, CUORE and other similar cryogenic experiments would benefit from continuous vibration monitoring, with witness channels placed directly on the MC plate or in other locations closer to the experimental volume.

In this article, we present a vibration monitoring setup using commercially available, cost-effective accelerometers that can be deployed 
at the MC stage of a DDR.
Using this setup, we monitored PT-induced vibrations at a broad range of stable temperatures, ranging from 8 mK up to 1 K, including an extended measurement at 8 mK. 
We discuss the experimental setup in Section~\ref{sec:setup}, including details on the DDR, selection of accelerometers and the data acquisition systems for temperature and accelerometer measurements.
The results of the vibration measurements are presented in Section~\ref{sec:results}.
Section~\ref{sec:radioassay} covers the setup and results of the radioassay on the accelerometers. 
In Section~\ref{sec:discussion}, we conclude with a discussion of the findings, potential improvements, and future prospects. 

\section{Experimental Setup}
\label{sec:setup}
All cryogenic vibration measurements in this study employed a triaxial configuration of three {Endevco 2271A piezoelectric accelerometers}~\cite{PCB_2271A} (weighing 27 g) mounted on the MC plate via a custom-machined oxygen-free electronic (OFE) grade copper mounting block.
The accelerometer outputs were fed into and amplified by a four-channel {PCB Piezotronics 482C64 conditioner}~\cite{PCB_482c64}, which features selectable charge-to-voltage conversion settings (0.1 mV/pC, 1 mV/pC, and 10 mV/pC), along with an additional incremental voltage gain setting between x0.1 and x200. 
The {2271A} has a charge sensitivity of 11.5 pC/g, specified over 1--8000 Hz frequency range and rated for operation down to 4K.
All data were acquired with the conditioner configured at a charge-to-voltage conversion factor of 10 mV/pC and a voltage gain of 42.74, resulting in a system gain of 4915 mV/g.
The overall gain was selected to prevent saturating the signal conditioner's dynamic range during typical cryostat vibration acquisition conditions.
The gains of all three {2271A} accelerometers were set identically to allow for relative comparisons between vibrations in each direction of the triaxial setup. 

\begin{figure}[htbp]
  \centering
    \includegraphics[width=0.8\textwidth,,trim={0 20cm 0cm 0},clip]{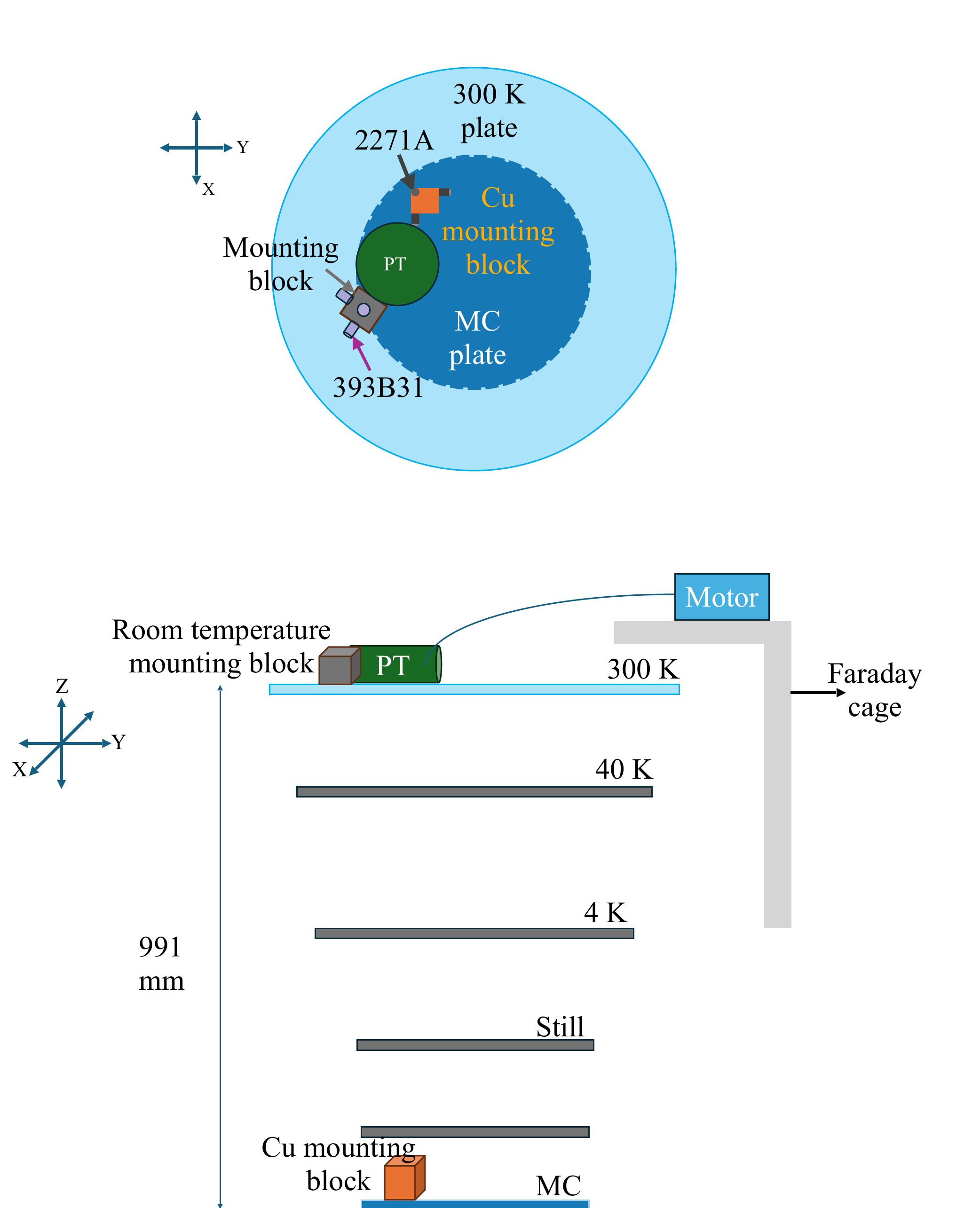}
  \caption{ (\emph{Top}): Schematic of the cryostat as seen from the top showing the placement of the mounting blocks. Also shown are their placements relative to the PT head.  The strongest impulse from PTs is approximately along the Y-axis on the schematic.
  Only the 300K and the MC plate are shown for the sake of clarity.
  (\emph{Bottom}): Lateral view of the measurement setup.}
  \label{fig:measurement_setup}
\end{figure}
\begin{figure}[ht!]
    \centering
    \includegraphics[width=0.65\textwidth]{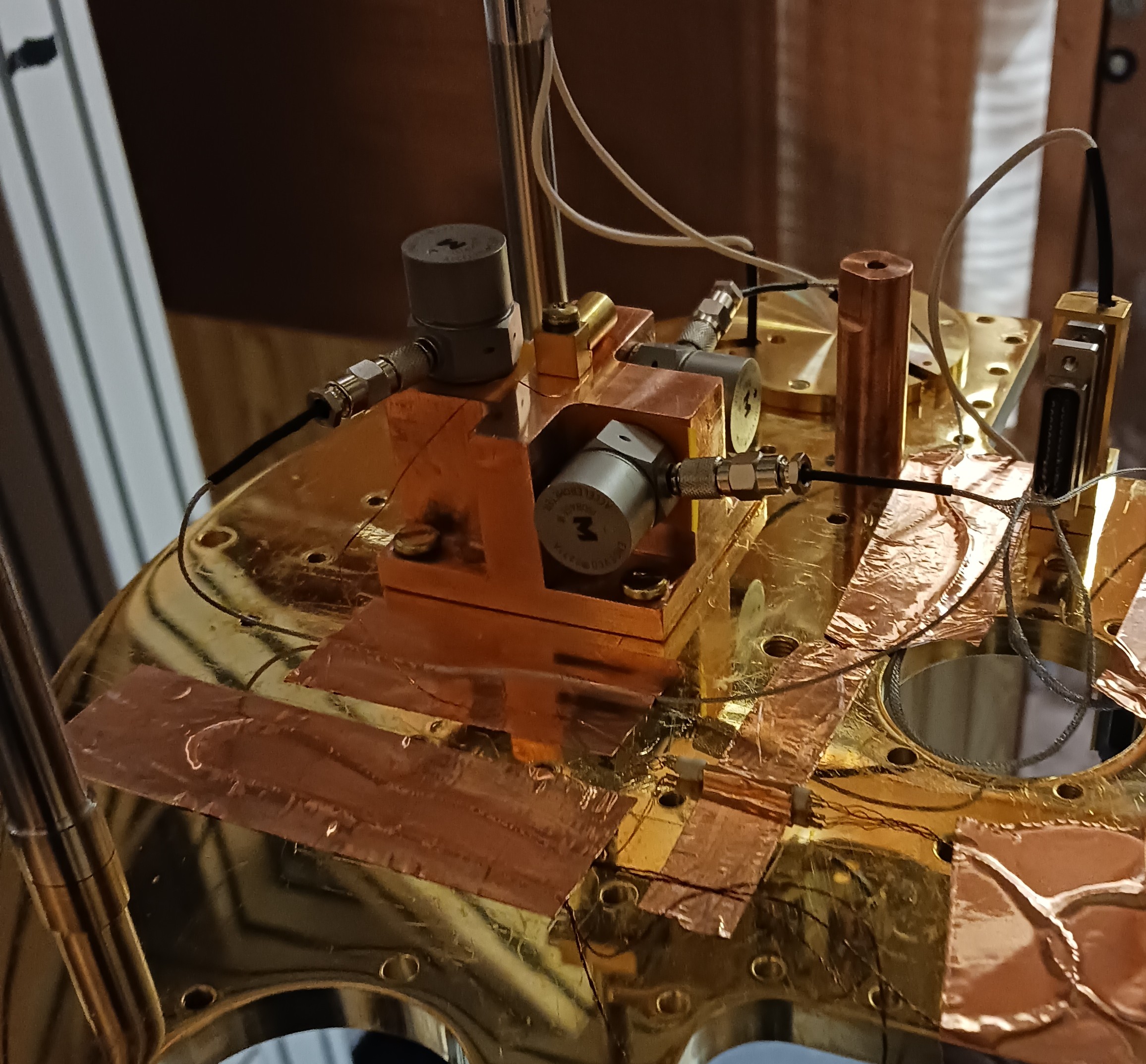}
    \caption{Photograph of the mounting block with all three {2271A} accelerometers mounted in a triaxial setup, as used for all the measurements. Also visible on top of the mounting block is the RuOx thermometer.}
    \label{fig:NbTi_setup}
\end{figure}
A separate triaxial setup with three {PCB Piezotronics 393B31} accelerometers~\cite{PCB_393b31} was used as a control to assess relative variations in the vibrational environment, with the accelerometers mounted on an aluminum block on the 300K plate in close proximity to the pulse tube cryocoolers. 
A four-channel {PCB Piezotronics 482C15 conditioner} ~\cite{PCB_482c15}, with selectable gain settings of x1, x10, and x100, was used to amplify the signals from the accelerometers.
All data were acquired with the conditioner gain setting configured at x1, which when combined with the rated sensitivity of {393B31} at 10 V/g (0.1--200 Hz), resulted in a total gain of 10000 mV/g.

The accelerometer-conditioner combinations for both the room temperature and cryogenic setups were chosen to maximize sensitivity at low frequencies~($\sim$1 Hz).
Since we did not perform an absolute calibration against a known vibration source, all accelerometer data are reported in arbitrary units and are intended for comparison only.
Figure~\ref{fig:measurement_setup} shows the placement of both sets of accelerometers, positioned so that one of the accelerometers approximately aligns with the direction of the strongest PT impulses.

\begin{figure}[ht!]
    \centering
    \includegraphics[width=1\textwidth]{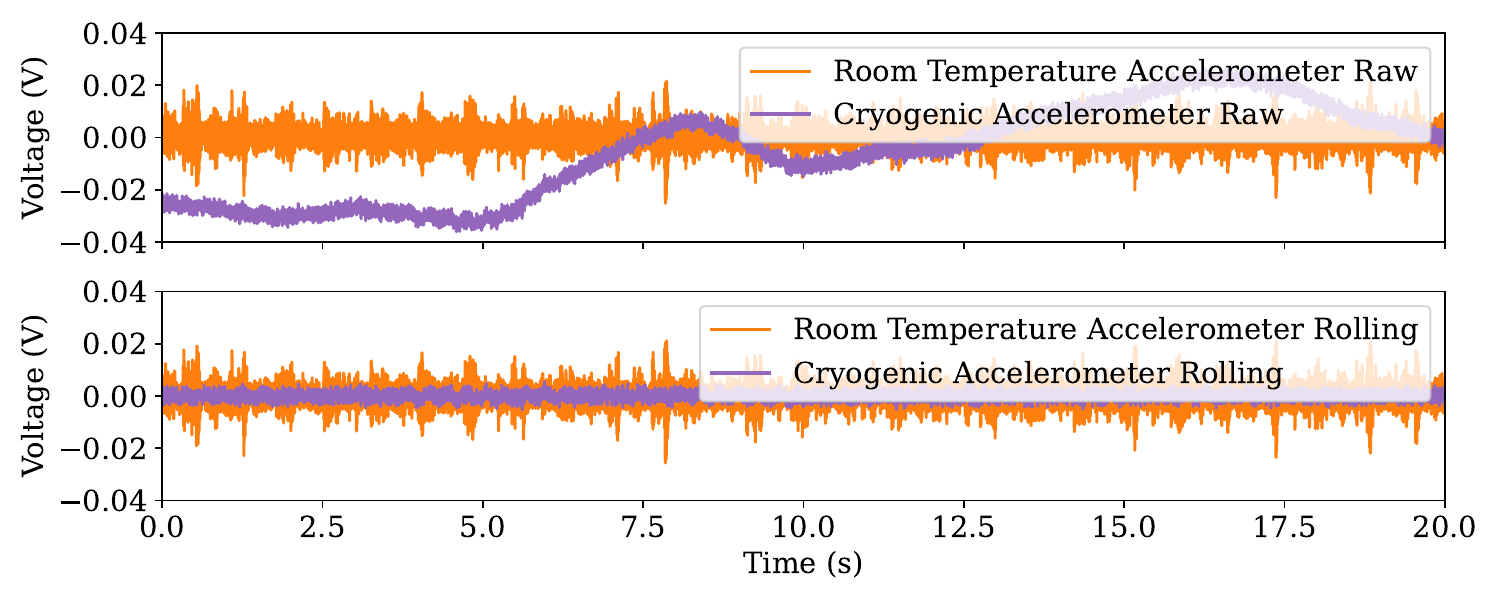}
    \caption{(\emph{Top}): Raw voltage time series data of a 393B31 accelerometer and a 2271A accelerometer, clearly showing a drift in the baseline of the {2271A} accelerometer. 
    (\emph{Bottom}): The same data smoothed by applying a 0.5 s rolling average subtraction.
    The baseline voltage drift was corrected for the 2271A accelerometers, while no noticeable variations were observed in the {393B31} accelerometers.}
    \label{fig:time_series}
\end{figure}
The signals from all six channels, routed through the respective conditioners, are fed to {National Instruments (NI) DAQ 9230 module}~\cite{PCB_9230} integrated with {NI cDAQ 9178 chassis}~\cite{PCB_9178} for digitization.
{9230} is a three-channel DAQ with 24-bit ADC resolution and a maximum sampling rate of 12.8 kS/s.
All data were acquired at a sampling rate of 1 kS/s.
An {NI LabVIEW} program was used for data acquisition, segmenting the data into 10-second intervals and recording them to disk.
An additional offline preprocessing step was used to merge the 10-second segments into a single long time series of the required duration.
To address slow drifts ($\sim$s-scale) observed in the~{2271A} data, we applied a 0.5 s rolling average subtraction, as shown in figure~\ref{fig:time_series}, to the entire time series for all datasets.

The measurements were carried out at a cryogenic facility located in Robeson Hall, Virginia Tech. 
The facility includes a {Bluefors LD400} DDR, equipped with a {Cryomech PT415} PT with a remote motor. 
A nearby {NESLAB HX750} air-cooled chiller cools the PT compressor and DDR gas handling system.  Vibrations from the gas handling system are decoupled from the cryostat by a bellows T-damper in the still pumping line.
To reduce electromagnetic interference, the refrigerator is housed inside a Faraday room.
The PT remote motor is mounted outside the Faraday room on an aluminum frame decoupled from the Faraday room, just above the room ceiling.
It is coupled via a $\sim$ 30 cm flex line to the PT cold head, which is mounted using a Bluefors-provided bellows to the 300 K plate of the cryostat. 
The 40K and 4K stages of the PT are thermally coupled to the 40K and 4K stages of the refrigerator via flexible copper braids.
The flexible, fine-wire structure of these copper braids, common in dilution refrigerators, is crucial for minimizing mechanical coupling while ensuring highly efficient thermal conduction. 
This reduced stiffness, compared to solid conductors, limits vibration transfer while maintaining high thermal conductivity.

The temperatures of the MC plate and the mounting block were monitored using two ruthenium oxide (RuOx) {Lakeshore RX-102B} thermometers.
These measurements were recorded using a {Lakeshore 372 AC} resistance bridge.
All temperature measurements in this study, unless explicitly stated, correspond to the thermometer attached to the MC plate.
The MC plate was also instrumented with a 120 $\Omega$ resistive heater to adjust the temperatures to desired values by applying configurable currents.

\section{Vibration Measurement Results}
\label{sec:results}
\begin{figure}[ht!]
    \centering
    \includegraphics[width=0.9\textwidth]{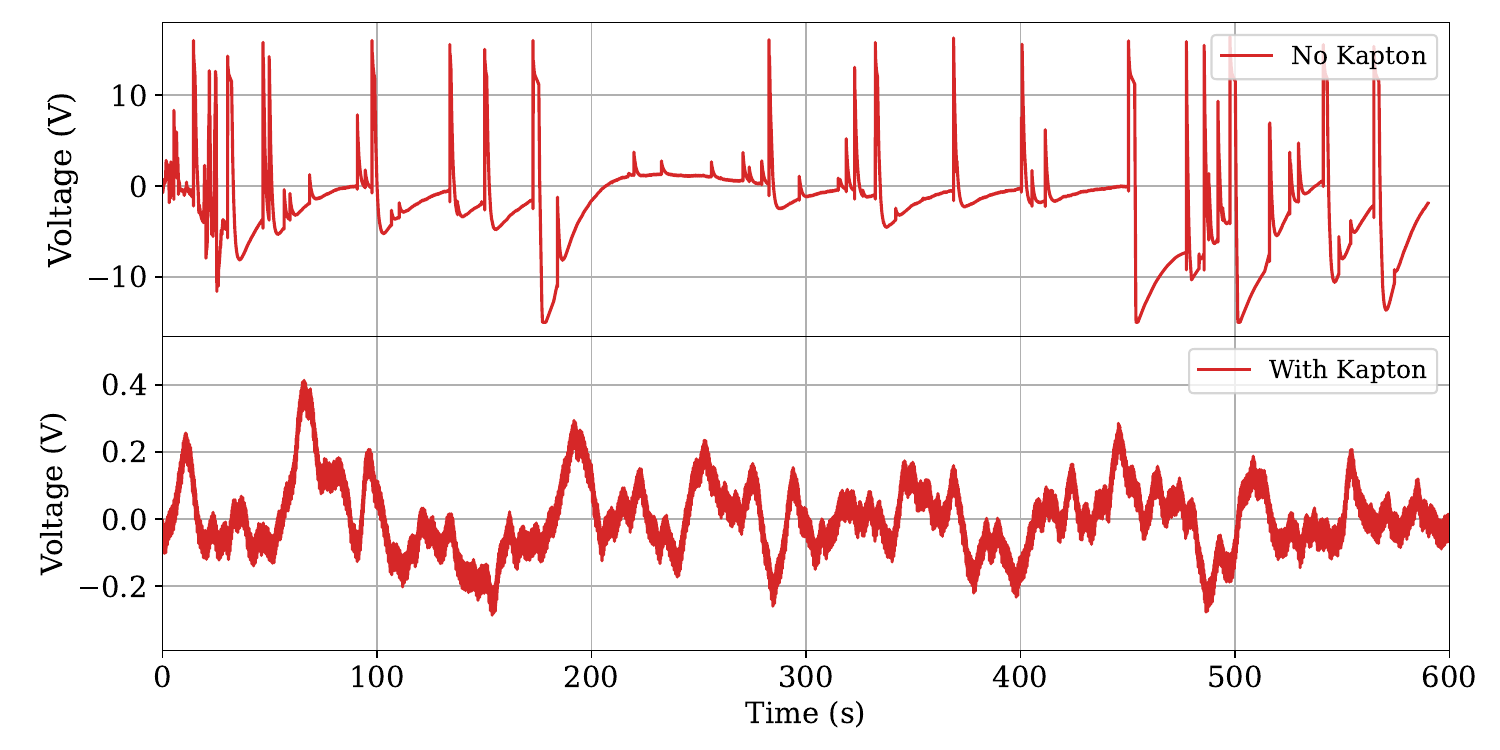}
    \caption{ (\emph{Top}): Time series of {2271A} accelerometer voltage on the MC plate with no Kapton tape electrical insulation. (\emph{Bottom}): Time series of {2271A} accelerometer voltage on the MC plate with Kapton tape on brass screws and the interface between mounting block and accelerometer. Notice the difference in voltage scales between the two figures.}
    \label{fig:kapton_vs_no-kapton}
\end{figure}
The data-taking began with the proprietary Endevco {3090C} cables provided with the cryogenic accelerometers.
The 10 ft long cables were thermalized at each stage by compressing a $\sim$15 cm segment between copper plates thermally connected to the stage.
These cables were connected to vacuum feedthroughs at the 300K plate, the other end of which were connected to the conditioners via BNC cables.
The cables carrying the signal from the feedthroughs to the DAQ were affixed to the exterior of the walls of the Faraday cage to minimize the effect of the cables' vibration on the accelerometer output. 
We found that the minimum temperature reached by the MC plate in this configuration was 60 mK, compared to 8 mK in the absence of load.
We attribute the additional heat load to the Endevco {3090C} cables~\cite{PCB_3090C} which were rated for a minimum temperature of 4 K.

We initially observed spurious electrical discharges manifesting as high voltage spikes.
To electrically isolate the body of the accelerometers from the copper mounting block, we used a thin layer of Kapton film between them.
Further, the body of the brass screws that were to secure the accelerometers to the mounting block, were electrically isolated using Kapton washers and sleeves.
These steps helped to diminish the spikes as shown in figure~\ref{fig:kapton_vs_no-kapton}.

\begin{figure}[ht!]
    \centering
    \includegraphics[width=0.9\textwidth]{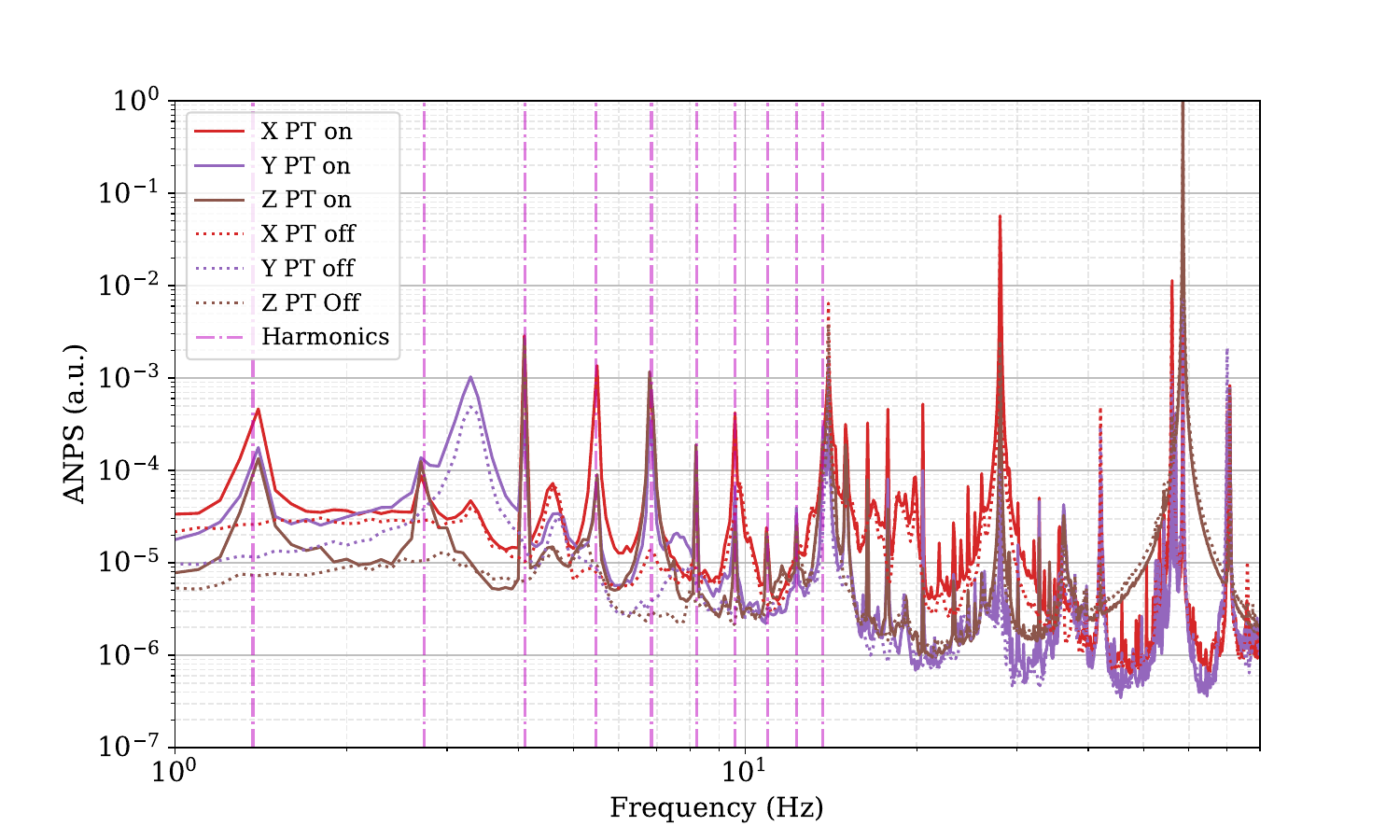}
    \caption{ANPS of 2271A accelerometers, after MC plate stabilized at 8 mK, when the PT was on~(solid lines) and turned off~(dotted lines). 
    Also shown are the 1.4 Hz base frequency of the PT and its next nine harmonics clearly illustrating the sensitivity of the accelerometers to vibrations. }
    \label{fig:low_temp}
\end{figure}
To minimize the heat load at the base temperature of the DDR, we switched the Endevco cables with a three-channel custom breakout cable consisting of short ($\sim$20 cm) segments of low-mass shielded NbTi cable (CMR 02-32-052) to couple each accelerometer output to the permanently installed low-mass readout wiring in the refrigerator.  
The permanently installed wiring terminates at an MDM25 connector on the MC plate and is composed of twisted pair NbTi from the MC plate to the 4K plate and twisted pair copper to the 300K flange.
One end of each channel was soldered to a Pasternack {PE44353} microdot connector to couple it to an accelerometer and the other end was soldered to the MDM connector. 
This change enabled the MC plate to achieve the cryostat's no-load base temperature of 8 mK.
The corresponding mounting block temperature stabilized at 15.2 mK.
Through the rest of the paper, we quote the temperature as measured on the MC plate.
\begin{figure}[ht!]
    \centering
    \includegraphics[width=0.9\textwidth]{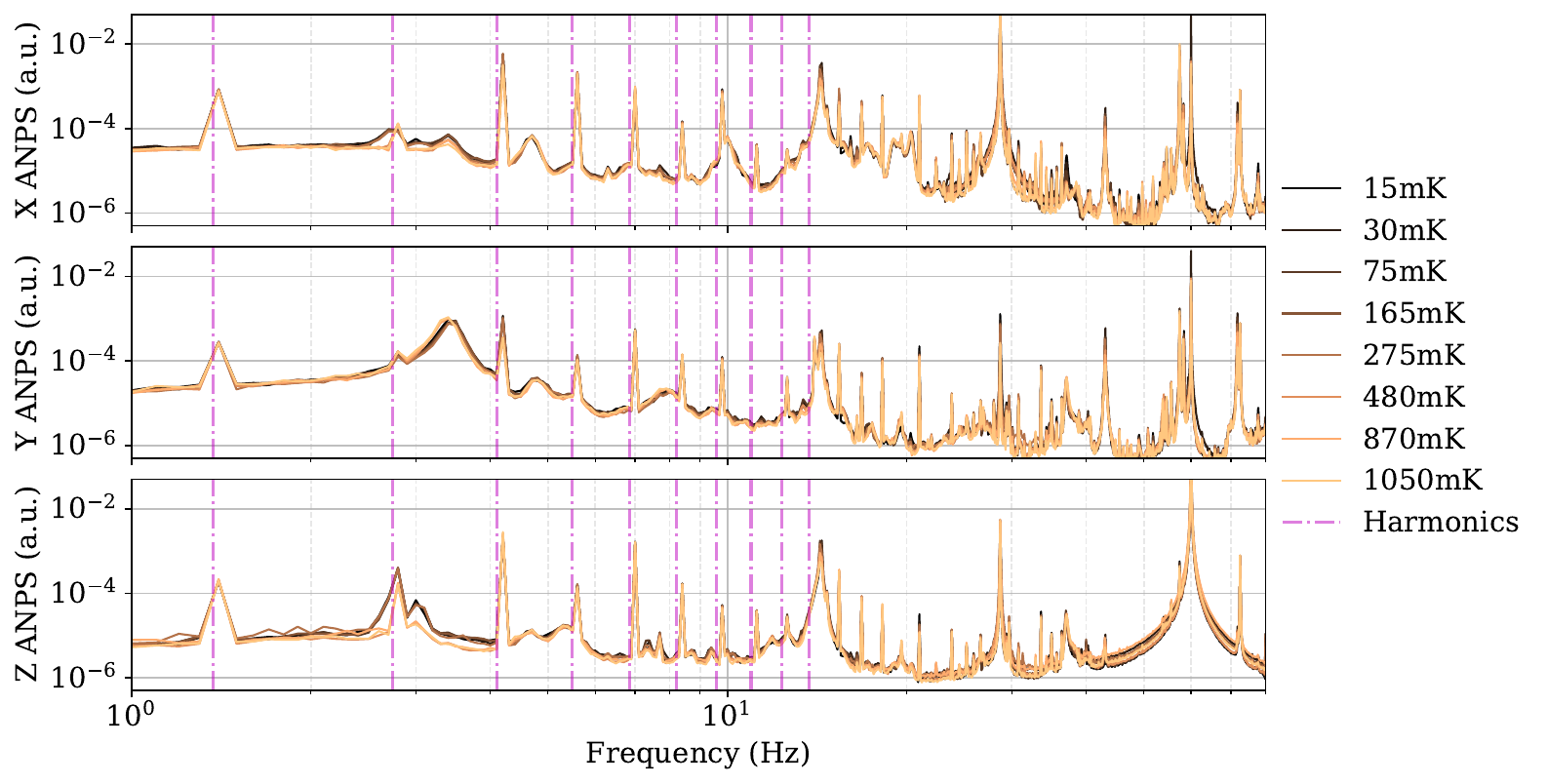}
    \caption{Stability of the response of all three 2271A accelerometers over MC plate temperatures ranging from 15 mK to 1050 mK, rounded to the nearest 5 mK. Each ANPS was generated using one hour of data.
    } 
    \label{fig:temp_stability}
\end{figure}

\begin{figure}[ht!]
    \centering
    \includegraphics[width=0.9\textwidth]{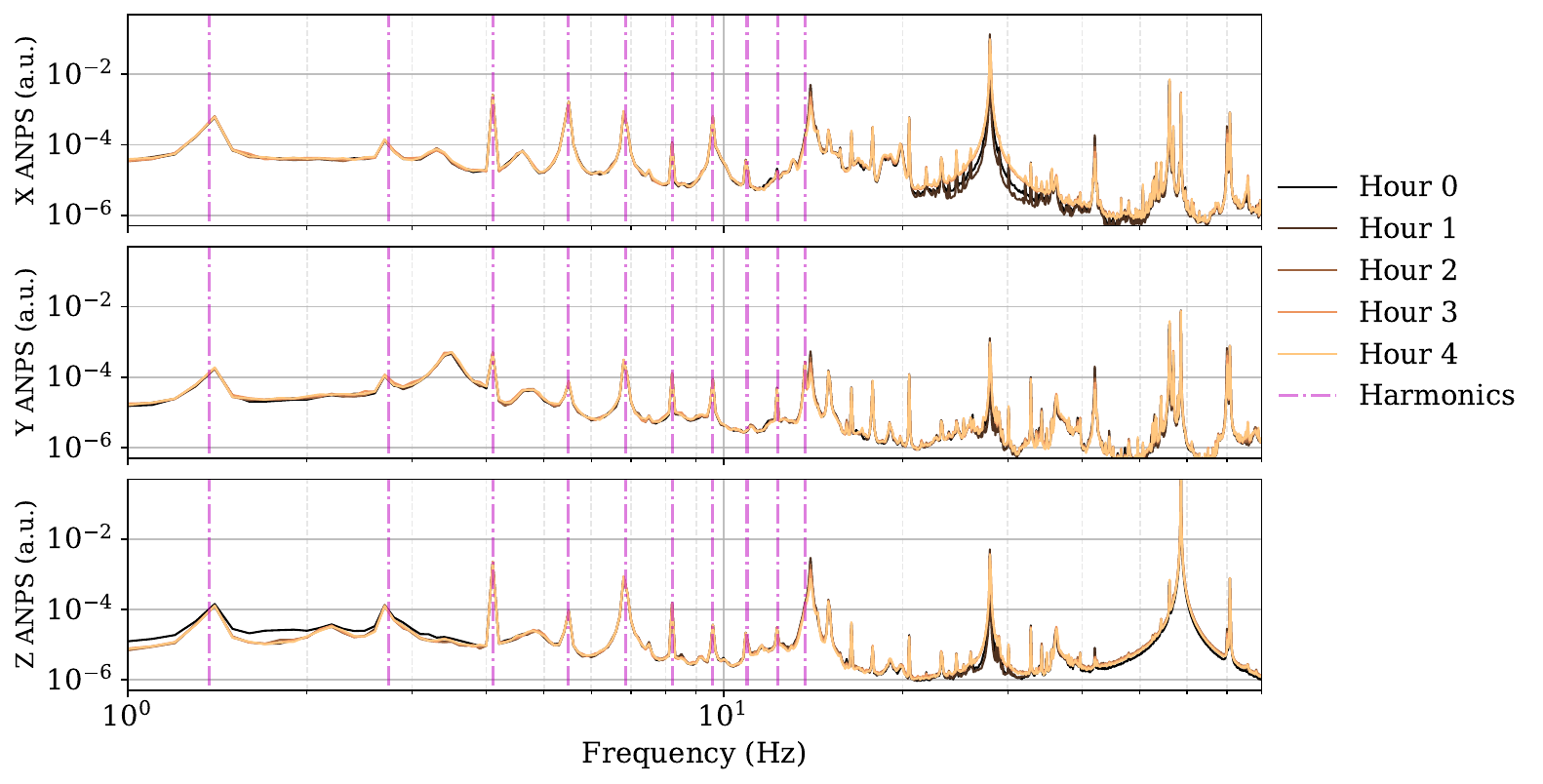}
    \caption{Stability of the response of the {2271A} accelerometers over five consecutive hours, separated into 1 hour segments. Average temperature of MC Plate during this data period was 8.0 mK. }
    \label{fig:time_stability_NbTi}
\end{figure}

Data from all six channels were processed by applying a rolling average subtraction, segmenting the data into 10 s windows, applying fast Fourier transforms (FFTs) to compute the power spectral density~(PSD), and averaging the results to obtain the average noise power spectrum~(ANPS).
The ANPS of the cryogenic accelerometers ({2271A}) at 8 mK are shown in figure~\ref{fig:low_temp}, clearly displaying PT harmonics, demonstrating that the accelerometers are sensitive to vibrations at 8 mK.
The PT harmonics observed in all three directions also demonstrate that the vibrations transmit to the MC plate both in the vertical and radial directions.

We also performed the noise measurement at the different temperatures by applying a heat load as discussed in Section~\ref{sec:setup} and letting the temperature equilibrate. 
Figure~\ref{fig:temp_stability} shows the ANPS of the accelerometers measured between 15 mK and 1050 mK, illustrating the stability of their performance across this temperature range.
Figure~\ref{fig:time_stability_NbTi} similarly demonstrates the time stability of the performance of the accelerometers at 8 mK measured over 5 hours. 

\begin{figure}[ht!]
    \centering
    \includegraphics[width=0.9\textwidth]{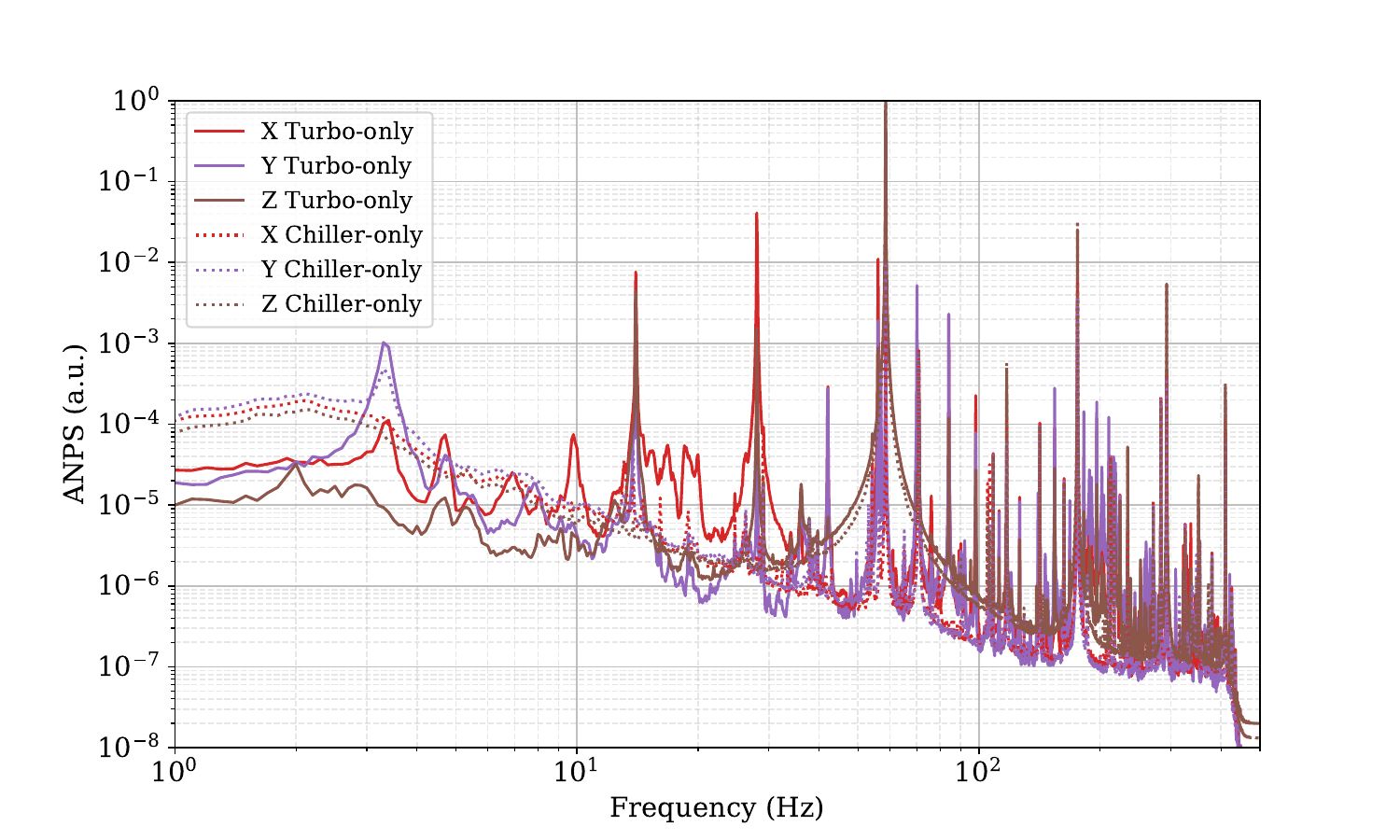}
    \caption{ {2271A} accelerometer response to cryogenic equipment. Solid lines show data collected when only the turbo pump is on. 
    Dotted lines show data collected when only the air cooled chiller is on. 
    The turbo data collection was kept short to ensure safe vacuum levels for the device operation.}
    \label{fig:Cryo_ChillerVsTurbo}
\end{figure}

We collected two datasets to isolate the vibrational noise contributions from the turbo pump and the air-cooled chiller.
The chiller-only data were collected for 2 hours by letting the MC reach base temperature, turning off PTs and turbo pump.
Similarly, turbo-only data were collected for 45 minutes by letting the MC reach base temperature, turning off PTs and chiller. 
Figure~\ref{fig:Cryo_ChillerVsTurbo} shows the chiller-only and turbo-only datasets illustrating that the turbo pump contributes several low frequency noise peaks. 
The turbo pump operates at a frequency of approximately 800 Hz. 
However, the interconnected assembly---which encompasses the dilution refrigerator, Faraday cage, support frame, and vacuum lines---is expected to manifest a complex spectrum of resonance modes, potentially including those at low frequencies. 
This finding aligns with observations documented in the relevant literature~\cite{10.1063/1.1416110}.

\section{Radioassay Results}
\label{sec:radioassay}
Measuring the radioactivity of devices installed in close proximity to the experimental volume is crucial for assessing their viability, especially for use in rare event searches.
Long-lived radioisotopes such as $^{40}$K, $^{238}$U and $^{232}$Th are of particular concern in low-background experiments.
The three accelerometers were assayed at Virginia Tech with a high purity germanium (HPGe) detector for approximately two weeks and the spectrum obtained was compared to a background spectrum (i.e., no sample present) also collected over two weeks. The detector is an Ortec Model GEM30 LLB-GEM-HJ HPGe gamma-ray spectrometer with an integrated cryostat with a relative efficiency of 35\%, energy resolution of 1.85 keV FWHM at 1332 keV, and is shielded on all sides by approximately 10 cm of lead. 
The detection efficiency for the accelerometer sample geometry was estimated with a GEANT4-based \cite{Geant} simulation implementing the nominal HPGe detector geometry with the accelerometers approximated as stainless steel cylinders and isotropic sources of $^{238}$U, $^{40}$K and $^{232}$Th confined to the accelerometer volumes. 

To validate the simulation, the efficiency was measured with point sources: $^{133}$Ba, $^{152}$Eu, $^{137}$Cs, $^{60}$Co and $^{22}$Na, which have an activity certified to 5\% uncertainty by the vendor~\cite{SpectrumTechniques}, and compared to simulations. Between 200 keV and 1410 keV, the ratio of the measured detection efficiency and detection efficiency extracted from simulation was found to be $0.711\pm 0.036$ independent of energy. We apply this correction factor to detection efficiencies extracted from the simulation of the accelerometer samples. As a further cross check,  a 30.2~g sample of K$_2$SO$_4$ powder with activity certified to 1.1\%~\cite{IAEA-RGK-1} was measured and simulated. The ratio between the measured detection efficiency and that extracted from the simulation was found to be $0.726\pm 0.015$. At 121 keV, we find the measured efficiency is 55\% of the efficiency extracted from simulation. At low energy, uncertainty on the detector geometry and possible presence of detector dead layer have a larger impact, and so we apply an additional systematic error on the efficiency below 200 keV. 
\begin{figure}[h!]
\centering
   \includegraphics[width=0.49\textwidth]{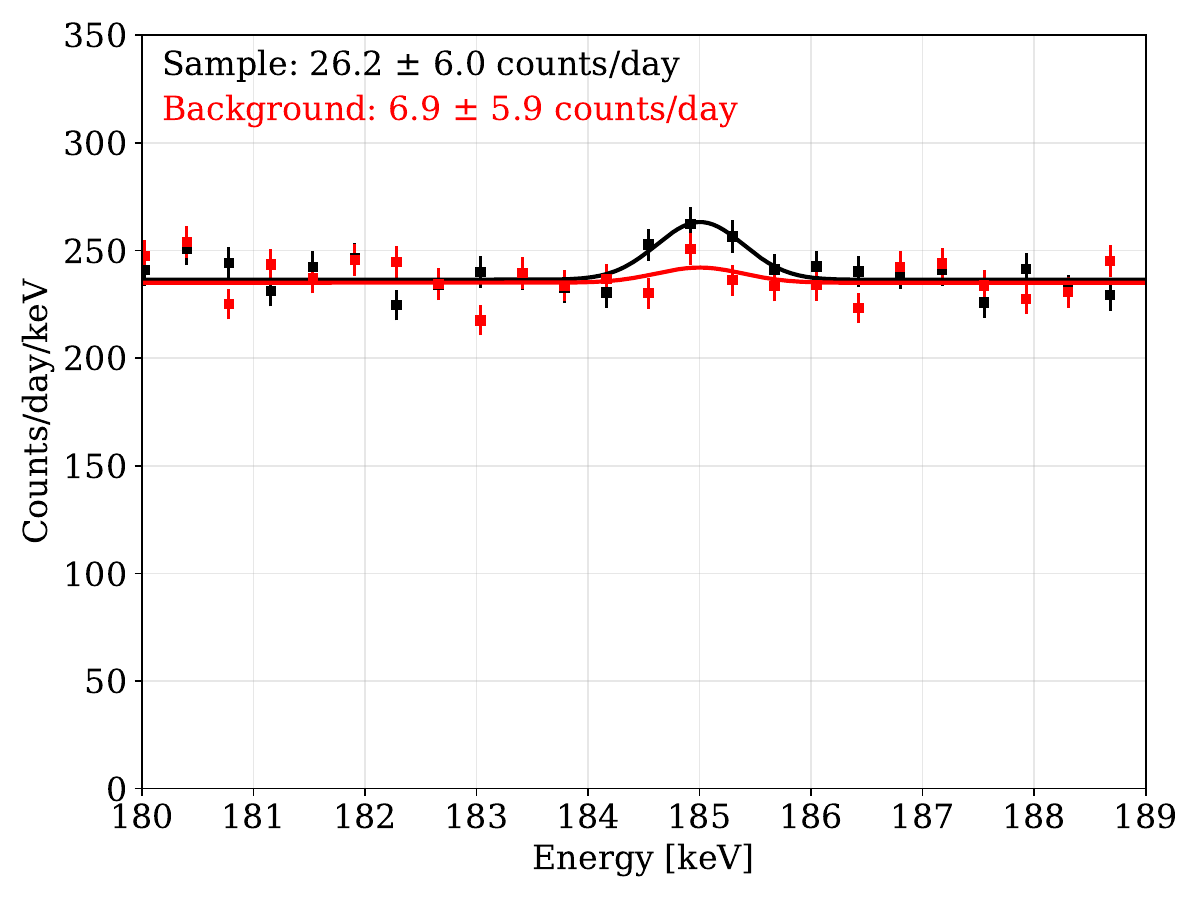} 
   \includegraphics[width=0.49\textwidth]{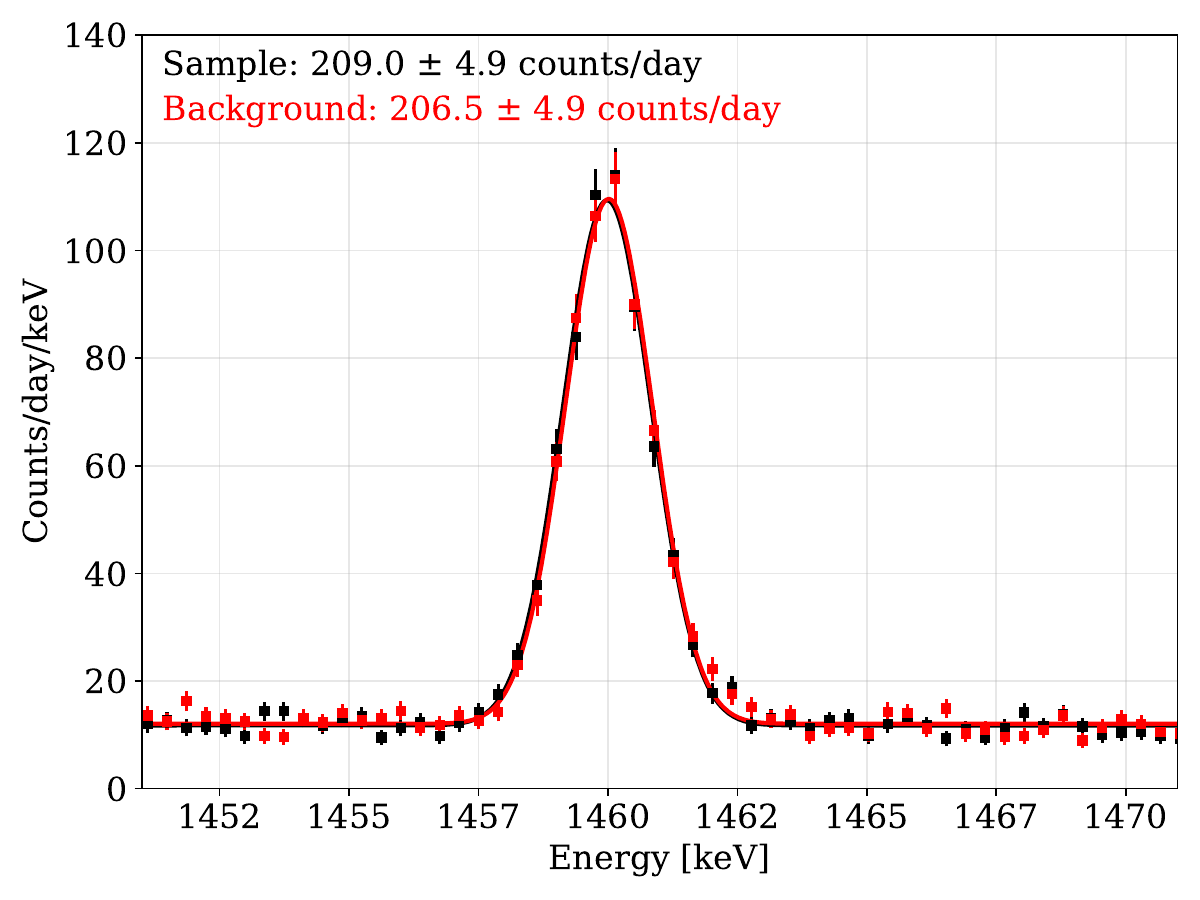} 
  \includegraphics[width=0.50\textwidth]{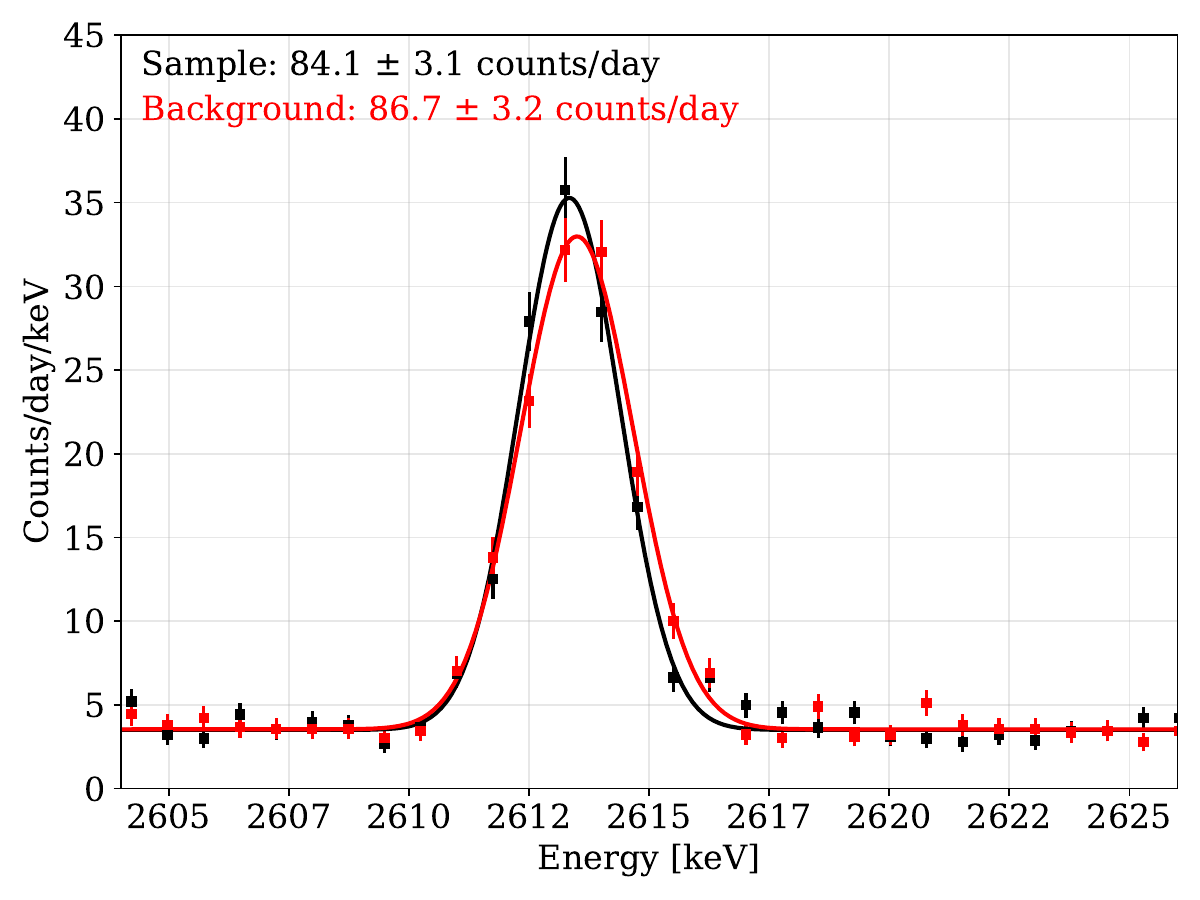} 
 \caption{Results of the HPGe counting analysis for the region of interest for $^{226}$Ra near 186 keV (\emph{top left}), $^{40}$K near 1460 keV (\emph{top right}), and $^{208}$Tl near 2614 keV (\emph{bottom}). The black histograms are the data collected with the sample, the red histograms are the data collected with no sample present. The black and red lines are the best-fit models assuming a gaussian gamma-peak and flat background. }
\label{fig:radioassay}
 \end{figure}

The results of the HPGe counting are summarized in Table~\ref{tab:radioassay} and shown in figure~\ref{fig:radioassay}.
The region of interest for each gamma line was modeled with a gaussian peak and flat background to determine the gamma yield with and without sample. We see no evidence of excess counts from $^{40}$K or $^{232}$Th from the sample. 
We do observe an excess of counts associated with the $^{238}$U chain. 
Since the $^{222}$Rn background in the counting setup can vary depending on airflow conditions in the lab, we rely on $^{226}$Ra line at $186$ keV to probe $^{238}$U contamination. 
We find the gamma yield in the ROI around 186~keV can depend on the range chosen for the fit due to the low peak-to-background ratio in that region. 
As an alternative, a simple counting between 183 keV and 188 keV yields $1274\pm 10 $ and $1235\pm 10$ counts per day with and without the sample, respectively.
Under the assumption that the $^{238}$U decay chain is in secular equilibrium, this corresponds to an activity of ($2.0\pm 0.8$)Bq/kg.
However, given that both the $^{238}$U and $^{232}$Th decay chains may be out of secular equilibrium, our future radioassay plans include a more sensitive measurement at an underground facility to determine the true activity without equilibrium assumptions.

\begin{table}[]
\begin{tabular}{|c|c|c|c|c|c|c|}
\hline
Parent & Gamma  & Bkg & Sample& Excess& $\epsilon$ & Activity\\
\cline{2-7} 
&(keV) & \multicolumn{3}{c|}{(counts/day)} & \%& Bq/kg\\
\hline
$^{238}$U & $^{226}$Ra (186) & $6.9\pm 5.9$      &     $26.2\pm6.0$        & $19.3\pm 8.4$        & $0.33^{+0.02}_{-0.08}$ & $0.9\pm 0.4$\\
\hline
$^{40}$K& $^{40}$K (1461)   &  $206.5\pm4.9$     &     $209.0\pm 4.9$      &   $<13$      &  $0.051\pm 0.003$ & <3.8\\
\hline
$^{232}$Th& $^{208}$Tl (2615) & $86.7 \pm  3.2$  &      $84.1\pm3.1$   & $<5.8$ & $0.088\pm 0.004$ & <1.0\\ 
\hline
\end{tabular}
\caption{Summary of radioassay results. The reported efficiency, $\epsilon$, is the estimated absolute detection efficiency for the specified gamma ray per decay of the parent source. Upper limits are at 90\% C.L., where the integral over the distribution for the excess parameter was restricted to non-negative values. The final column is the estimated specific activity of the specified parent source in the accelerometer samples. }
\label{tab:radioassay}
\end{table}

\section{Discussion}
\label{sec:discussion}

This work demonstrates the use of cryogenic accelerometers to perform vibrational noise measurements at the coldest stage in a DDR across a broad frequency range.
It paves the way for various applications, including continuous monitoring of the vibrational environment in DDRs.

The ability to coherently measure noise and assess vibration transmission across different locations opens up a range of applications.
As an example, it enables applications such as denoising leveraging continuous, real-time measurements that facilitate ongoing noise assessment and mitigation in vibration-sensitive experiments.
As a use case, we consider one of the tri-axial 393B31 accelerometers as the device of interest and the 2271A accelerometers as the \emph{witness} sensors and generate the coherence and transfer function~(TF) to characterize the vibration transmission across the various DDR stages.

\begin{figure}[ht!]
    \centering
    \includegraphics[width=\textwidth]{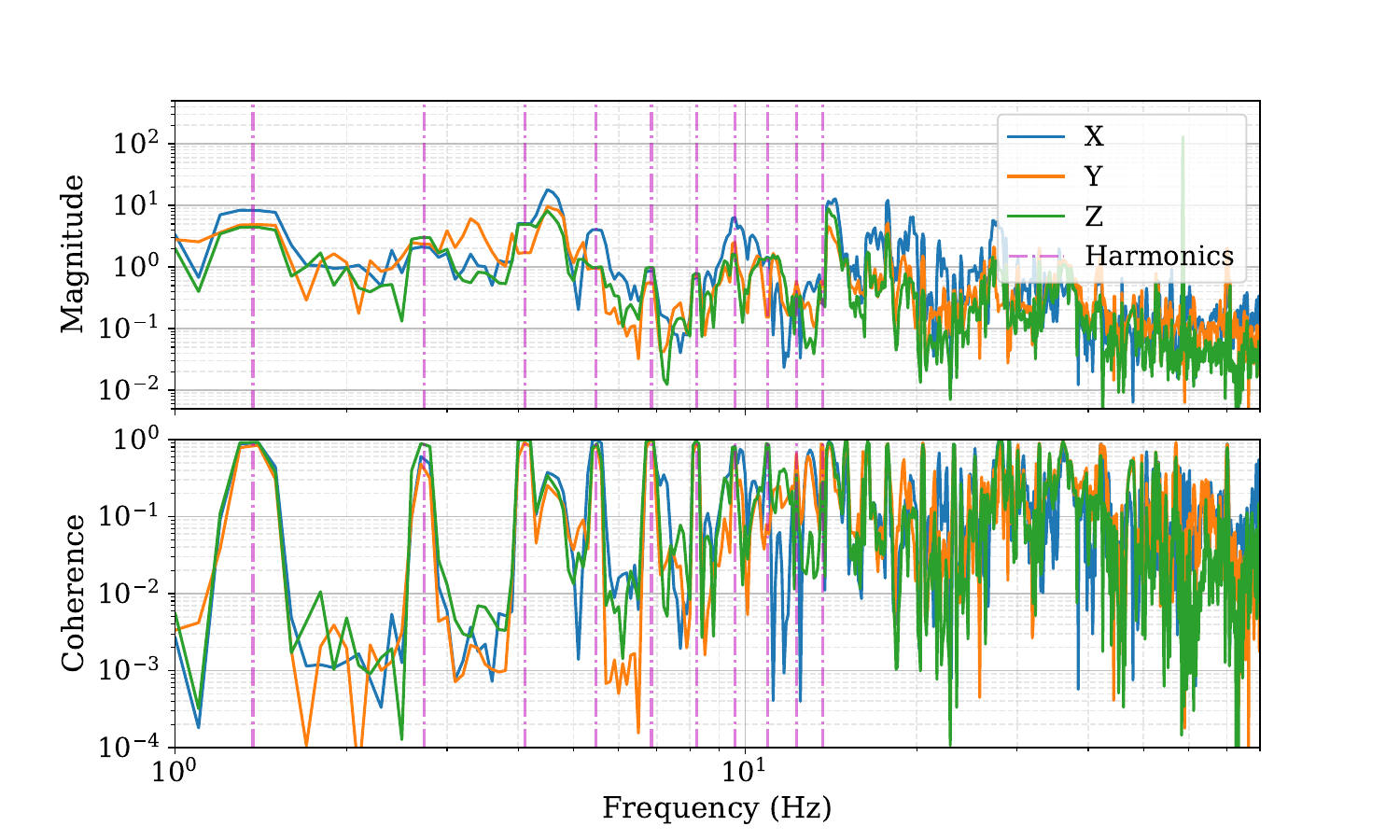}
    \caption{ (\emph{Top}): Transfer function magnitude between {2271A} accelerometers in respective directions for 1 hour of data against the {393B31} accelerometer in Y direction. The Y direction was selected as it was the direction with the most sensitivity to PT induced vibrations for the {393B31} accelerometers.
    (\emph{Bottom}): Coherence between identical sets of channels as described above.}
    
    \label{fig:TF}
\end{figure}

Following the procedure developed in Ref.~\cite{Vetter:2023fas}, we begin by defining the cross-spectral density between any two devices $i$ and $j$ as the time-averaged products of the discrete Fourier transforms (DFT) $\mathcal{F}_{i}(f)$ and $\mathcal{F}_{j}(f)$ of their time streams, respectively as
\begin{align}\label{eq:CSD_Aux_to_Aux}
    G_{i,j}(f) = \frac{2}{T}\langle \mathcal{F}_i^*(f) \cdot \mathcal{F}_j(f) \rangle 
\end{align}
where $T$ is the duration of time windows used for averaging. 
$G_{i,j}(f)$ is a matrix of size $N$, where $N$ is the number of DFT bins.
The level of correlation between devices for all frequencies can be quantified using $G_{i,{k}}$, dropping $f$, as
\begin{equation}\label{eq:coherence}
    C_{i,k} = \frac{|G_{i,k}|^2}{G_{i,i} G_{k,k}}
\end{equation}
The TF between device $i$ and device $k$ can then be computed as
\begin{equation}\label{eq:TransferFunction}
  H_{i,{k}} = \sum_{j =1}^{n} G_{i,j}^{-1}G_{j,{k}},
\end{equation}
where $n$ is the number of witness sensors.
$H_{i,{k}}$ is a complex $n\times N$ matrix and defines the relationship between device $i$ and device $k$ across all frequencies.
Figure~\ref{fig:TF} shows the coherence and the TF between
cryogenic accelerometers and $y$-aligned room temperature accelerometer.

To use a series of witness sensors $i$ as a way to identify correlations and perform denoising for the data $O_{k}$ from the sensor $k$, the TF, $H_{i,{k}}$, could be used as follows:
\begin{equation}\label{eq:denoised_signal}
   D_{k} = O_{k} - \sum_{i =1}^{n} H^{T}_{i,{k}} A_{i}
\end{equation}
where $D_{k}$ is the denoised signal. As long as there are correlations between the witness sensors and the detector, this algorithm performs reasonably well. 

This technique was successfully demonstrated~\cite{CUORE:2024ikf}, and continues to be used, by the CUORE collaboration to denoise their bolometer signals with a series of accelerometers, seismometers, and antennas acting as witness sensors. 
Stronger noise correlations between witness sensors and the detector lead to a better denoising performance.
For detectors coupled to the lowest stage of a DDR with complex inter-stage couplings---such as CUORE---deploying witness sensors at multiple stages could enhance correlations.

A critical consideration in deploying cryogenic accelerometers in DDRs is the amount of heat they dissipate at mK temperatures. 
Excessive heat load at this temperature can make it impossible for DDRs to reach their base temperature, and consequently compromise their performance.
To estimate the heat load associated with the setups using Endevco 3090C cables and the custom NbTi cables, we used the measured temperature-power curve of the DDR. 
This curve was obtained previously with no payload attached to MC by applying known heat loads to the resistive heater discussed in Section~\ref{sec:setup}.
The power required for the DDR MC to stabilize at 60 mK, the base temperature for the Endevco 3090C cable setup, was calculated in this way to be 119 $\mu$W. 
The NbTi cable setup, by maintaining the MC base temperature at 8 mK, contributes negligible heat load.
For context, the cooling power of the Blue Fors LD400 employed in this work is 15 $\mu$W at 20 mK.
This heat load estimation underscores the need for a careful thermal management when integrating accelerometers into ultra-low temperature cryogenic environments. 

In addition to thermal considerations, the intrinsic radioactivity of the accelerometers poses a potential limitation for experiments searching for rare events, such as neutrinoless double beta decay.
As discussed in Section~\ref{sec:radioassay}, HPGe spectroscopy measurements indicate that most $\gamma$ lines arising from long-lived isotopes of the uranium and thorium decay chains are consistent with ambient background levels, with the one exception being 186 keV $\gamma$ line associated with $^{226}$Ra decay arising from the $^{238}$U chain, with a corresponding activity of (2.0 $\pm$ 0.8)Bq/kg.
Although the observed activity is not particularly high and remains inconsequential for most cryogenic applications, incorporation of optimized passive shielding might be needed for low background experiments.
As part of future investigations, a higher-sensitivity radioassay is being considered at an underground counting facility.

These results represent a critical advancement in vibrational sensing using commercial accelerometers at ultra-low temperatures, yet they highlight the challenges posed by noise in practical deployments.
The primary sources of noise stem from electronic interference and environmental signals within the accelerometer-cable assembly, with triboelectric effects emerging as the likely major contributor.
While custom NbTi cables were used to minimize heat load, Endevco cables were found to exhibit substantially lower noise at mK temperatures.
This indicates that lower noise levels could be achievable beyond what is demonstrated in these results.

Mounting blocks represent another key area for noise reduction enhancement.
While copper block was chosen for mounting the 2271A accelerometers, 
preliminary results using a different cryostat show that anodized aluminum blocks exhibit lower noise drift down to 3~K compared to copper, suggesting a viable alternative.
Thus, commercially available anodized aluminum mounting blocks for the 2271A accelerometers will be assessed.

Immediate efforts beyond this work will focus on quantifying noise sources and optimizing cable and mounting configurations. 
Concurrently, calibration using precision shakers will be explored to establish reliable sensitivity metrics and improve signal-to-noise ratios for real-world applications.
Longer term work includes accelerometer placement tests to systematically optimize the vibrational noise detection in complex vibrational environments.

\acknowledgments
This work is supported by the University of Pittsburgh and the US Department of Energy (DOE) Office of Science, Office of Nuclear Physics under Contract Award Nos. DE-SC0020423 and DE-SC0011091.
This work is supported by the Piano Nazionale di Ripresa e Resilienza (PNRR), Missione 4 “Istruzione e Ricerca” - Componente 2, Investimento 1.1, “Fondo per il Programma Nazionale di Ricerca e Progetti di Rilevante Interesse Nazionale (PRIN)”, project code 2022KRKM2X, Title: “Ultra-low vibrations thermal switch for Pulse Tube cryocoolers”, CUP I53D23001270006.
The authors thank CUORE and CUPID collaborators for productive discussions, technical support, and comments on the paper draft.
The authors acknowledge Advanced Research Computing at Virginia Tech for providing computational resources and technical support that have contributed to the results reported within this paper. \href{https://arc.vt.edu/}{https://arc.vt.edu/}.
\bibliographystyle{JHEP}
\bibliography{inputs/biblio}
\end{document}